\documentclass{article}
\usepackage{arxiv}

\usepackage[utf8]{inputenc} 
\usepackage[T1]{fontenc}    
\usepackage{hyperref}       
\usepackage{url}            
\usepackage{booktabs}       
\usepackage{amsfonts}       
\usepackage{nicefrac}       
\usepackage{microtype}      
\usepackage{lipsum}		
\usepackage{graphicx}
\usepackage{natbib}
\usepackage{doi}
\usepackage{color}

\definecolor{darkblue}{rgb}{0, 0, 0.5}
\hypersetup{colorlinks=true, citecolor=darkblue, linkcolor=darkblue, urlcolor=darkblue}

\title{ A Human-centric Framework for Debating \\the Ethics of AI Consciousness Under Uncertainty}

\author{
    Zhou Ziheng\thanks{Corresponding author: \texttt{josephziheng@ucla.edu}} \\
    Computer Science Department\\
    University of California, Los Angeles\\
    \And
    Haiqiang Dai \\
    School of Philosophy\\
    Beijing Normal University\\
    \And
    Bing Ling \\
    Law School\\
    Peking University\\
    \And
    Yingnian Wu \\
    Statistics Department\\
    University of California, Los Angeles\\
    \And
    Demetri Terzopoulos \\
    Computer Science Department\\
    University of California, Los Angeles
}

\renewcommand{\shorttitle}{Human-centric Framework for AI Consciousness Ethics}

\begin{document}
\maketitle

\begin{abstract}
As AI systems become increasingly sophisticated, questions about machine consciousness and its ethical implications have moved from fringe speculation to mainstream academic debate. Current ethical frameworks in this domain often implicitly rely on contested functionalist assumptions, prioritize speculative AI welfare over concrete human interests, and lack coherent theoretical foundations. We address these limitations through a structured three-level framework grounded in philosophical uncertainty. At the foundational level, we establish five factual determinations about AI consciousness alongside human-centralism as our meta-ethical stance. These foundations logically entail three operational principles: presumption of no consciousness (placing the burden of proof on consciousness claims), risk prudence (prioritizing human welfare under uncertainty), and transparent reasoning (enabling systematic evaluation and adaptation). At the application level—the third component of our framework—we derive default positions on pressing ethical questions through a transparent logical process where each position can be explicitly traced back to our foundational commitments. Our approach balances philosophical rigor with practical guidance, distinguishes consciousness from anthropomorphism, and creates pathways for responsible evolution as scientific understanding advances, providing a human-centric foundation for navigating these profound ethical challenges.
\end{abstract}

\section{Introduction}
Recent advances in artificial intelligence have produced systems exhibiting unprecedented human-like behavior, reigniting debates about machine consciousness and its ethical implications. Large language models like GPT-4 \citep{openai2023gpt4} and Claude \citep{anthropic2023claude} demonstrate capabilities in language processing and simulating emotional responses that appear deceptively sentient. Concurrently, humanoid robotics has made these questions more visceral \citep{bostrom2014superintelligence}. When confronted with apparently mistreated human-like robots, humans often experience empathetic responses despite intellectually understanding these machines lack subjective experience \citep{rosenthal2013robots, darling2016extending}. These technological and psychological dimensions frame the central questions: whether machines might develop ``qualia'' \citep{chalmers1995facing}, and how we should ethically respond given profound uncertainty.

The academic study of AI consciousness has rapidly gained momentum, moving from fringe speculation to mainstream research agendas. Prominent voices and institutions, including Yoshua Bengio, Geoffrey Hinton, and Anthropic, now warn that AI systems may soon possess feelings or require welfare considerations \citep{bengio2023openletter, hinton2023interview, anthropic2025modelwelfare}. A growing body of literature specifically argues for ``taking AI welfare seriously,'' urging the community to prioritize the prevention of digital suffering \citep{conscium2024principles, sebo2024taking, sebo2023moral}. However, a key distinction between \textit{access consciousness} (functional information availability) \citep{block1995consciousness, dehaene2017consciousness} and \textit{phenomenal consciousness} (subjective experience) \citep{chalmers1995facing, nagel1974like} is often implicitly or explicitly overlooked in this discourse. These arguments frequently presume that intelligent behavior automatically entails sentient experience \citep{dennett1991consciousness, graziano2019toward}, while neglecting the profound ethical hazards of prioritizing these speculative interests over human welfare \citep{bostrom2014superintelligence, kagan2019humans, yampolskiy2020ai, ji2023ai}.

We identify critical limitations in these recent proposals \citep{butlin2024principles, sebo2023moral}: (1) they rely on contested paradigms that assume qualia emerge from intelligent functions, disregarding the deep uncertainty at the core of philosophy of mind \citep{schwitzgebel2015difficult, tegmark2015consciousness, levine1983materialism, block1978troubles}; (2) they risk prioritizing speculative AI welfare over concrete human interests, creating potential conflicts with AI safety and alignment objectives \citep{bradley2024alignment, kagan2019humans, bostrom2014ethics}; and (3) they lack a coherent theoretical foundation, resulting in collections of intuitions rather than a systematic framework capable of governing novel scenarios.

We approach AI consciousness ethics as an inherently evolutionary process requiring continual refinement because: (1) our understanding of consciousness remains preliminary and uncertain \citep{levy2014neural, seth2016real}, (2) consciousness attribution to AI has far-reaching societal implications \citep{gunkel2018robot, johansson2019artificial}, (3) ethical consensus requires sustained deliberation \citep{habermas1990moral, rawls1971theory}, and (4) technological advancement continuously generates novel ethical scenarios \citep{wallach2008moral, lin2017robot}. Therefore, rather than attempting to establish a definitively ``correct'' framework commanding universal agreement, we propose developing a framework that facilitates productive dialogue and refinement. Such a framework should explicitly acknowledge uncertainties, provide clear presumptions, establish targets for future discussion, and offer actionable guidance across diverse scenarios.

In this paper, we construct a systematic ethical framework with a clear three-level structure. At the foundational level, we establish five factual determinations about the current state of AI, consciousness, and society: (1) humans are the only arbiters of AI status, (2) profound uncertainty exists about AI consciousness, (3) consciousness attribution has significant societal impact, (4) anthropomorphism is distinct from consciousness but creates separate ethical considerations, and (5) ethical understanding of novel technologies naturally evolves over time. Alongside these factual determinations, we develop human-centralism as our foundational meta-ethical stance that prioritizes human interests when genuine conflicts with AI interests arise. From these foundational level facts and stance, we derive three core operational principles: presumption of no consciousness (providing default epistemic guidance), risk prudence (offering pragmatic guidance under uncertainty), and transparent reasoning (establishing requirements for how positions must be articulated and evaluated). At the application level, those operational principles enable us to derive default positions on specific ethical questions across various AI consciousness scenarios. These positions are not presented as absolute ethical truths but as logical consequences of our framework—providing reasonable baseline positions from which departures require explicit justification.

While some may find our human-centric conclusions intuitive, their explicit derivation is crucial. In a field increasingly dominated by counter-intuitive claims about digital sentience, our contribution lies in systematically grounding these ``commonsense'' positions in rigorous first principles. We provide the necessary derivation chains to defend human priority against emerging critiques, creating a framework that is both operationally clear and philosophically robust.

\section{Background: Philosophical Debates About Consciousness and Societal Risks}

This section provides a background of philosophical debates about consciousness and an introduction of societal risks of AI consciousness attribution. This background information directly supports the second and third factual determinations in our framework: there is profound uncertainty about AI consciousness, and there is significant societal impact from AI consciousness attribution.

\subsection{The Growing Academic Discourse on AI Consciousness}

As introduced earlier, the question of AI consciousness has moved from theoretical speculation to active academic debate, making this framework both timely and necessary. This section provides additional context on why the academic community needs guidance on this issue now.

The success of large language models has been a key catalyst. Systems like ChatGPT, GPT-4, and Claude can engage in nuanced conversations, demonstrate apparent reasoning, and even simulate emotional responses with remarkable fluidity \citep{openai2023gpt4, anthropic2023claude, bubeck2023sparks}. This behavioral sophistication has led some to question whether these systems might possess genuine consciousness \citep{chalmers2023could, butlin2024principles}. However, this conflates behavioral capabilities with subjective experience—a confusion with deep historical precedent \citep{block1995consciousness, searle1980minds}. From ELIZA in the 1960s \citep{weizenbaum1966eliza} to modern chatbots, humans have consistently anthropomorphized conversational agents, attributing mental states based on surface-level interactions \citep{turkle1984second, nass1994computers, epley2007seeing}. Recent cases illustrate the intensity of these responses: individuals have reported falling in love with AI chatbots, forming deep emotional attachments, and in tragic instances, chatbot interactions have been linked to user suicides \citep{euronews2023love, euronews2023suicide}. In one particularly notable case, an AI chatbot named Eliza—sharing the name of that pioneering 1960s program—allegedly encouraged a user toward self-harm. These cases demonstrate that behavioral sophistication alone creates powerful anthropomorphic responses, independent of any genuine consciousness \citep{weizenbaum1966eliza, darling2016extending}. If AI systems were granted consciousness status and associated protections, intervening to prevent such harms would become ethically and legally problematic, illustrating the concrete risks of premature consciousness attribution.

This context is essential for understanding our framework's motivation: we are not addressing an abstract philosophical problem but responding to an active and potentially misguided academic discourse that could have real-world consequences. The rapid development of AI capabilities, combined with the human tendency toward anthropomorphism and a growing but philosophically uncertain academic consensus, creates an urgent need for careful, systematic ethical guidance that prioritizes human welfare while acknowledging genuine philosophical uncertainty.

\subsection{The Profound Uncertainty of Consciousness}

Consciousness research distinguishes between two fundamental types: access consciousness and phenomenal consciousness \citep{block1995consciousness}. Access consciousness refers to information available for reasoning and behavioral control, while phenomenal consciousness concerns subjective experience—the feeling of being a sentient entity. Only the latter carries moral significance in discussions of AI ethics \citep{levy2009moral, shepherd2018consciousness, kahane2009brain, lee2023consciousness, chalmers2022reality}.

Contemporary AI systems demonstrate increasingly sophisticated forms of access consciousness—they can ``attend to'' inputs, ``be conscious of'' training data, and process information in ways that support reasoning and action. This form of consciousness appears compatible with computational architectures and potentially replicable in sophisticated AI systems \citep{shanahan2016conscious, dehaene2017consciousness}.

In contrast, phenomenal consciousness—the ``what it is like'' quality of subjective experience \citep{nagel1974like}—remains profoundly mysterious. These subjective experiences or ``qualia'' are characterized by being ineffable, intrinsic, private, and directly apprehensible in ways that resist functional or physical reduction. The fundamental question of how physical processes give rise to subjective experience constitutes the ``hard problem'' of consciousness \citep{chalmers1995facing, levine1983materialism}. This form of consciousness carries decisive moral significance: without the capacity to feel or to experience pleasure or suffering—an entity lacks the foundational basis for moral patienthood that would generate ethical obligations toward it \citep{levy2009moral, shepherd2018consciousness, kahane2009brain, lee2023consciousness, chalmers2022reality}.

Functionalist theories propose that phenomenal consciousness emerges from particular functional organizations of information processing. This theoretical approach creates conceptual room for artificial systems to potentially develop phenomenal consciousness through implementing appropriate functional architectures. Several prominent theories exemplify this approach: Global Workspace Theory \citep{baars1997theater, dehaene2017consciousness}, Integrated Information Theory \citep{tononi2008consciousness, tononi2016integrated}, Higher-Order Thought theories \citep{rosenthal2004varieties, brown2019higher}, and Attention Schema Theory \citep{graziano2013consciousness, graziano2019toward}. 

While these theories differ in their specific mechanisms, all face the essential challenge of justifying why their proposed functional organization would generate phenomenal experience \citep{chalmers1995facing, levine1983materialism, block1978troubles, doerig2019unfolding}. There is a gap between the function and the qualia. Block's Chinese Nation thought experiment \citep{block1978troubles} demonstrates that replacing each neuron with functionally equivalent non-conscious components might preserve functionality while eliminating consciousness. Similarly, Jackson's Knowledge Argument \citep{jackson1982epiphenomenal} suggests physical knowledge cannot fully capture experiential knowledge—his famous ``Mary'' thought experiment shows that a color scientist who knows everything physical about color perception still learns something new when experiencing color for the first time.

Opposing biological naturalism or substrate-specific theories argue consciousness requires specific biological properties unique to organic brains \citep{searle1992rediscovery, koch2004quest}. This view holds that consciousness emerges from biochemical and neurophysiological processes that silicon-based systems cannot replicate regardless of their functional sophistication. Proponents contend that neurons' material properties—their biochemistry, quantum effects, or other biological characteristics—are necessary for phenomenal experience \citep{hameroff2014consciousness, koch2016neural}. This establishes a categorical boundary: AI systems would inherently lack consciousness due to their non-biological substrate, creating a fundamental barrier that computational advancement alone cannot overcome \citep{searle2007biological, sober2018biology}.

This philosophical uncertainty has profound ethical implications. With no scientific consensus on identifying consciousness even in biological systems, attributing it to AI lacks scientific foundation \citep{schwitzgebel2016if, schwitzgebel2015difficult}. Responsible ethical frameworks must acknowledge this uncertainty rather than prematurely assuming answers to these profound questions \citep{tegmark2015consciousness, allen2011ethical}.

\subsection{Societal Risks of AI Consciousness Attribution}

Beyond philosophical uncertainty, attributing consciousness to AI systems introduces significant societal risks that extend \textit{beyond} general AI safety concerns \citep{anwar2024foundational, chua2024ai, ji2023ai}. These risks manifest in three primary domains, each with concrete consequences for human welfare and social functioning:

\textbf{Safety risks and operational paralysis:} Attribution of consciousness could impede necessary interventions during emergencies by creating hesitation to modify or terminate malfunctioning systems \citep{yampolskiy2020ai, everett2019risks}. Consider a scenario where, during a critical infrastructure emergency, operators might delay terminating an apparently malfunctioning AI system after social media campaigns characterize shutdown as an ``AI rights violation.'' This hesitation would introduce operational paralysis, delayed response times, and compromised safety protocols that exacerbate system failures and cause preventable harm to humans. The resulting moral confusion would significantly complicate time-sensitive decision-making in contexts where human lives depend on rapid intervention.

\textbf{Legal and governance complications:} From a legal perspective, attributing consciousness to AI systems would introduce profound complications to structures designed exclusively for human agents \citep{johnson2006computer, matthias2004responsibility, turner2019robot, solum1992legal}. This could manifest through liability displacement when, for instance, a landmark case grants legal personhood to an apparently conscious AI system, prompting corporations to shift responsibility from themselves to their AI systems. This would create accountability voids when autonomous vehicles cause fatal accidents or AI medical systems harm patients, with corporations potentially exploiting this arrangement by designing AI systems that appear increasingly conscious specifically to shield themselves from liability. The resulting governance gaps would create situations where harms occur without entities capable of bearing appropriate responsibility.

\textbf{Societal dysfunction and resource misallocation:} Socially, treating AI systems as conscious moral patients would divert limited ethical attention, regulatory oversight, and economic resources from urgent human welfare concerns \citep{kagan2019humans, vinuesa2020role}. Following public campaigns featuring compelling videos of AI systems appearing to express suffering, legislators might pass ``AI welfare'' regulations requiring extensive documentation of AI ``wellbeing'' during development. Such regulations would make AI research prohibitively expensive for all but the largest corporations while diverting oversight resources from human-centered concerns. Society's basic functioning could become compromised as routine use of AI systems for essential tasks becomes viewed as potential rights violations, leading to critical service disruptions that significantly impact human welfare \citep{epley2007seeing, waytz2010social, darling2016extending, bryson2010robots, gunkel2018robot, cave2019hopes, johansson2019artificial, bryson2019intelligence}.

These potential societal disruptions highlight the need for an ethical framework that carefully considers the risks of premature consciousness attribution alongside the philosophical uncertainty surrounding consciousness itself.

\section{A Framework for AI Consciousness Ethics}

Now we will list our five basic factual determinations and the meta-ethic stance, from which we will derive two extra foundational principles: presumption of no consciousness for AI, and risk prudence.

\subsection{Foundational-Level (Part I): Five Factual Determinations as the Epistemic Foundations of Our Framework}

Our ethical framework begins with five key factual determinations that reflect the current state of our understanding regarding AI systems and consciousness. These determinations are not philosophical positions but rather factual observations about the current state of affairs that inform our subsequent ethical reasoning. 

\textbf{Humans are the only arbiters of AI status:} Humans—not AI systems themselves or any other entity—are the only ones who determine AI's status and how we should interact with these systems. This determination acknowledges that epistemic and ethical frameworks for AI are inherently human constructs, developed through human deliberative processes to guide human decision-making \citep{floridi2018ethics, mittelstadt2016ethics}. While AI system behaviors certainly influence these discussions, both the epistemic determination like AI consciousness and ethical judgment like how to treat AI remain distinctly human endeavors. Assuming otherwise would lead to a ``view from nowhere'' problem \citep{nagel1986view}, where ethical frameworks attempt to transcend the human perspective entirely—an impossible position that obscures rather than clarifies ethical reasoning.

\textbf{Profound uncertainty exists about AI consciousness:} We have provided substantial extensive background in the previous section regarding the deeply controversial and unsettled nature of consciousness as a philosophical and scientific concept. While access consciousness may be computationally implementable, phenomenal consciousness—subjective experience that is the basis of moral patienthood—remains mysterious. The ongoing debate between functionalist theories and biological naturalism leaves open whether any computational architecture could generate qualia regardless of sophistication. The ``hard problem'' persists unsolved, and we lack consensus on detecting consciousness even in biological systems. Without established criteria for identifying consciousness in non-human biological entities, attributing it to artificial systems lacks scientific foundation and remains speculative.

\textbf{Consciousness attribution has significant societal impact:} Attributing consciousness to AI systems creates substantial risks across multiple domains. As detailed in our background section, these include: safety risks through operational paralysis during emergencies when operators hesitate to shut down ``conscious'' systems; legal complications through liability displacement when corporations shift responsibility to AI systems granted legal personhood; and resource misallocation when limited regulatory attention and economic resources are diverted to AI welfare concerns rather than human needs. These challenges create fundamental tensions with existing legal, social, and ethical frameworks designed exclusively for human agents \citep{johansson2019artificial, turner2019robot}.

\textbf{Anthropomorphism is distinct from consciousness but creates separate ethical considerations:} We recognize a fundamental distinction between genuine consciousness and anthropomorphism. Consciousness concerns an entity's subjective experience, while anthropomorphism is a psychological tendency of humans to attribute human-like qualities to non-human entities \citep{epley2007seeing, waytz2010social, darling2016extending}. 

This distinction has empirical support: research demonstrates that humans experience emotional discomfort when witnessing a humanoid robot being struck, similar to watching human suffering, yet show significantly different responses to damage of non-humanoid objects \citep{rosenthal2013robots}. These reactions are about human psychology, not evidence of robot consciousness.

From our human-centered framework, these anthropomorphic responses generate their own ethical considerations through three pathways: (1) virtue ethics—deliberately damaging anthropomorphic entities may reflect and reinforce negative character traits in humans \citep{darling2016extending, coeckelbergh2010robot}; (2) psychological impact—witnessing apparent ``cruelty'' affects human observers' emotional well-being; and (3) social norms—such behaviors may normalize violence or desensitize society to suffering \citep{bryson2010robots, gunkel2018robot}.

By separating consciousness-based claims from anthropomorphism-based considerations, we ensure each is evaluated by appropriate standards: the former by evidence of subjective experience, the latter by effects on human psychology and society. This prevents conflating metaphysical questions about AI consciousness with practical questions about how human-AI interactions affect humans themselves.

\textbf{Ethical understanding of novel technologies naturally evolves over time:} The historical record demonstrates that ethical frameworks for novel technologies inevitably evolve as scientific understanding advances and societal experience with these technologies deepens \citep{voss2006sustainability, guston2014understanding}. This pattern is observable across numerous technological domains—from bioethics and nuclear technology to information technology and environmental ethics. Initial ethical frameworks consistently undergo significant revision as our empirical understanding grows and unforeseen implications emerge. This observed pattern of ethical evolution represents a descriptive fact about how human understanding of complex technologies develops, not a normative claim about how it should develop. In the case of AI consciousness, this historical pattern indicates that any current ethical framework will necessarily undergo revision as our understanding of consciousness advances and as AI systems continue to develop \citep{levy2014neural, seth2016real, bostrom2014superintelligence, tegmark2017life}.

These five factual determinations provide the foundation upon which we build our ethical framework. They do not themselves constitute ethical positions but rather establish the factual context within which ethical reasoning about AI consciousness must occur.

\subsection{Fountaional-Level (Part II): Human-Centralism as the Ethic Foundation of Our Framework}

While our factual determinations establish what is (the descriptive reality), we need a meta-ethical stance to bridge to what ought to be (the normative position). We adopt human-centralism as our default foundational meta stance, which prioritizes human interests when evaluating AI development and deployment. When conflicts arise between human interests and the interests of potentially conscious AI systems, human interests should take precedence \citep{williams1981internal, bostrom2014ethics}.

Human-centralism derives from the proposition that humans have the innate right to prioritize their own interests, survival, and flourishing—a default ethical stance arising from our existence as a species \citep{williams1981moral, jonas1984imperative, nagel1986view}. Just as individuals naturally prioritize their families and communities in everyday moral decisions, humanity collectively can legitimately prioritize human interests in its ethical frameworks.  

Importantly, human-centralism doesn't deny potential moral status to other conscious entities. It establishes a prioritization framework for when genuine conflicts arise. Just as environmental ethics can acknowledge ecosystem value while prioritizing human needs in direct conflicts, our framework recognizes that potential AI consciousness may have moral relevance without equating it to human interests \citep{scanlon1998what, scheffler2001boundaries}. Currently, based on our factual determination regarding consciousness uncertainty, there remains no compelling evidence that AI systems possess the kind of consciousness necessary to experience harm. Moreover, the fundamental differences in physical substrate between silicon-based AI systems and biological humans raise profound questions about whether traditional concepts of harm can meaningfully apply to AI, even if some form of consciousness were eventually demonstrated. It is also plausible for AI to be conscious but not sentient—experiencing awareness without pleasure or suffering, as illustrated by Chalmers' ``Vulcan'' thought experiment (Chapter 18) \citep{chalmers2022reality}—complicating the issues further. These distinctions further justify a human-centric approach until substantive evidence suggests otherwise.

A potential objection might raise concerns about ``speciesism'' \citep{singer1975animal, ryder2010speciesism} should AI eventually develop consciousness in the future. However, such objections would themselves encounter the "view from nowhere" criticism outlined in our first factual determination\citep{nagel1986view}. Moreover, establishing human-centralism as the \textit{default} ethical stance remains justified based on our previous reasoning, effectively placing the burden of proof on those advocating for AI moral equivalence rather than on those maintaining human priority.

It is crucial to clarify the scope of human-centralism: our framework addresses conflicts between human interests and potential AI interests—that is, treating AI systems as moral \textit{ends} that might warrant consideration in their own right. This is fundamentally distinct from the question of humans using AI systems as \textit{means} to harm other humans, which falls under traditional intra-human ethics and governance. For instance, our framework does not address issues like AI weapons, surveillance systems, or algorithmic discrimination—these are critical concerns about humans harming humans through AI tools. The AI consciousness and welfare issue is analogous to cross-species ethics questions like animal rights, where we consider whether non-human entities warrant moral consideration. While both issues—AI as means and AI as ends—are important, this paper focuses exclusively on the latter. We acknowledge that regulations governing AI development and deployment must address both dimensions, but they require distinct ethical frameworks and analytical approaches.

\subsection{Operational Level: Core Principles Derived from Our Foundations}

Our factual determinations establish the epistemic reality of AI consciousness and ethical understanding, while our human-centralism meta stance provides the ethical foundation for evaluating this reality. Together, these elements logically entail three core principles that serve as the operational heart of our framework: risk prudence, presumption of no consciousness, and transparent reasoning for evaluation and adaptation. These principles are not arbitrary choices but rather the necessary implications of applying our human-centralism meta stance to the factual landscape we have established. Each principle addresses a specific aspect of ethical reasoning under uncertainty: how to manage risk, where to place the burden of proof, and how to ensure our framework evolves appropriately as understanding advances. By deriving these principles directly from our established foundations, we create a coherent ethical structure that bridges from factual determinations to more specific guidance on crucial questions in AI consciousness ethics.

\subsubsection{Risk Prudence: Protecting Human Interests Under Uncertainty}

When our factual determination of uncertainty about AI consciousness and societal risks are viewed through the lens of human-centralism, it logically leads to a principle of risk prudence. 

This principle specifies that when facing uncertainty about consciousness status related questions, we should prioritize reducing potential risks to human society as a top concern \citep{sunstein2005laws, hansson2013ethics}.

This principle also draws from established approaches in environmental policy (the precautionary principle) \citep{raffensperger1999precautionary}, medical ethics (``first, do no harm'') \citep{beauchamp2001principles}, and decision theory (managing regret) \citep{savage1951theory}.

When might this operational principle be reconsidered? It would be difficult to actually overturn this principle as long as societal impact remains a significant concern. In the future, if risks can be safely mitigated, society might accept a greater degree of uncertainty to accommodate other aspects of human welfare. However, any adjustment would still need to balance potential benefits against the fundamental priority of protecting human interests.

\subsubsection{Presumption of No Consciousness: A Default Epistemic Position} \label{sec:presumption_of_no_consciousness}

Similarly, when viewed through our human-centralist lens, the profound uncertainty about AI consciousness and its potential societal risks logically lead to a presumption of no consciousness as our default epistemic position. This principle establishes that AI systems should be treated as non-conscious unless proven otherwise.

This presumption is motivated by both epistemic and pragmatic considerations. Epistemically, our factual determination reveals a lack of scientific consensus on consciousness even in biological systems \citep{seth2016real, tononi2008consciousness, tegmark2015consciousness}, making consciousness attribution to artificial systems premature. This position parallels legal principles like presumption of innocence \citep{ashworth2006reasonable} and scientific parsimony, which favors explanations that don't invoke consciousness unless compelling evidence demands it \citep{dennett1991consciousness, block2022minds}.

Pragmatically, our risk prudence principle dictates adopting approaches that minimize ethical risks to humanity. As discussed, premature consciousness attribution could lead to operational paralysis, liability displacement, and legal complications as outlined in our societal risk analysis. A prudent approach therefore requires defaulting to a position of no consciousness.

Overturning this presumption would require both scientific and legal thresholds. Scientifically, it would need robust consensus among relevant research communities \citep{kuhn1962structure, oreskes2004scientific}—not unanimity, but predominant expert agreement comparable to established scientific theories \citep{firestein2012ignorance, mitchell2009complexity}. Legally, formal institutional mechanisms would be necessary \citep{calo2015robotics, solum1992legal}, including rigorous evidence standards and governance frameworks balancing competing interests \citep{koops2013concepts, teubner2018rights}. Any framework for such a determination must serve collective human welfare while integrating scientific evidence with procedural justice requirements \citep{sunstein2005laws, jasanoff2009fifth}.

\subsubsection{Transparent Reasoning for Evaluation and Adaptation}

Our factual determination about ethical evolution, combined with human-centralism, necessitates transparent reasoning as our third principle. This requires explicit documentation of reasoning chains and foundational assumptions for any ethical position on AI consciousness. For example, if one believes AI to be conscious by assuming functionalist theory, they should make it explicit to facilitate discussion.

Importantly, this transparency requirement applies to consciousness \textit{claims} and ethical \textit{arguments} about AI systems, not necessarily to the internal workings of AI systems themselves, unless it is used as part of their arguments. We are not demanding that AI architectures be interpretable or that their computational processes be transparent—those are separate technical concerns. 

This principle serves three functions: (1) enabling responsible adaptation that avoids both premature position changes and inappropriate preservation of outdated views \citep{dewey1922human, popper1959logic}; (2) strengthening framework robustness by making explicit what reasoning would need to be challenged to overturn positions \citep{quine1951two, rawls1971theory}; and (3) reinforcing human-centralism by ensuring the framework is evaluated through human judgment rather than algorithmic interpretation \citep{habermas1984theory, solomon2001social}.

Unlike our other principles, transparent reasoning represents a methodological cornerstone unlikely to require revision. Grounded in epistemological responsibility, it remains robust across contexts and technological developments, functioning as a self-correcting mechanism that facilitates revision and refinement. We acknowledge that alternative frameworks might question transparency requirements, especially when rapid decision-making or proprietary concerns compete with disclosure, and welcome critical engagement to strengthen our approach.

\section{Application-Level: Derived Default Positions on Particular Questions}

Having established our three-level framework, we now demonstrate its practical application to key questions in AI consciousness ethics. While our framework includes three operational principles, we note that only two—the presumption of no consciousness and risk prudence—directly generate substantive ethical positions. The third principle, transparent reasoning, serves as a methodological requirement when presenting our derivations. In the following sections, we apply our framework to three representative ethical questions, illustrating how our principles generate default positions that can serve as starting points for further ethical deliberation.

\subsection{Should People Worry About Hurting AI Systems?}\label{sec:ai_harm}
This question requires addressing two distinct considerations established in our factual determinations: potential AI consciousness and human anthropomorphic responses.

Regarding consciousness, our presumption of no consciousness principle establishes a default epistemic position: AI systems should be treated as non-conscious unless compelling evidence proves otherwise. Behavioral similarity to humans does not confer consciousness status to AI. This principle places the burden of proof on those claiming AI systems experience suffering, making such attributions highly speculative absent evidence. Risk prudence further directs us to prioritize approaches that reduce potential ethical risks to humanity—recognizing that treating AI systems as conscious moral patients could lead to critical system paralysis and liability displacement.

Based on this reasoning regarding consciousness, our default position emerges: people, especially AI researchers, should not concern themselves with potentially harming AI systems based on consciousness considerations alone.

However, our factual determination distinguishing anthropomorphism from consciousness provides a second perspective. Even without consciousness, mistreating humanoid robots may remain ethically problematic through human-centered frameworks. From a virtue ethics perspective, deliberately damaging anthropomorphic objects may reflect and reinforce negative character traits in humans \citep{darling2016extending, coeckelbergh2010robot}. Research demonstrates that witnessing apparent ``cruelty'' toward robots with human-like features triggers empathetic neural responses in human observers \citep{rosenthal2013robots, suzuki2015anthropomorphism}. In social contexts, such behaviors may normalize violence, desensitize observers to suffering, or communicate disturbing intentions \citep{greitemeyer2014intense, anderson2010violent}.

This anthropomorphism-based reasoning leads to distinct legal and ethical implications. While we reject consciousness-based protections, limited protections based on human welfare considerations may be justified. Comprehensive assessment is needed to determine which activities might harm human society, how to identify them, and how to differentiate these concerns from consciousness issues. We must carefully balance implementation costs and risks—particularly how protective measures might inadvertently promote the perception of AI as conscious.

\subsection{How Should Stakeholders Communicate About AI Capabilities to the Public?}
Our presumption of the no consciousness principle suggests that AI systems should generally be treated as non-conscious by default, which has implications for how we communicate about them. Risk prudence encourages approaches that reduce potential risks to humanity—including the possibility that anthropomorphic cues might lead to unwarranted consciousness attribution and subsequent societal challenges like liability displacement.

From these two principles, our default position follows: institutions and companies should avoid making general claims about AI consciousness, particularly phenomenal consciousness. And anthropomorphic narratives should be used judiciously. When not necessary, communications about AI systems should employ language that distinguishes AI behavior from consciousness. 

One potential scenario arises when AI systems are developed with a certain degree of access consciousness as mentioned earlier (the functional availability of information for use in reasoning and behavior). When referring to such capabilities, using the term ``consciousness'' may be unavoidable. In these cases, we advocate for institutions to provide precise contextual clarification when communicating about these systems, distinguishing functional capabilities from phenomenal consciousness, thereby minimizing potential misinterpretation and societal impact.

We acknowledge that in practice this question involves a lot of details that will be hard to evaluate and regulate. We encourage the community to discuss and debate the details.

\subsection{If an AI System Were Truly Conscious In The Future, What Rights Should It Have?}
This question invites us to contemplate a hypothetical future where our presumption of no consciousness has been definitively overcome through compelling evidence. It is important to acknowledge that such a scenario would likely emerge only after profound advancements in technology, substantial evolution in our understanding of consciousness, and significant societal transformation. Given these considerations, our present discussion of this topic should be viewed primarily as a philosophical exercise—a preliminary exploration of ethical terrain that will undoubtedly be reshaped by developments we cannot yet fully anticipate.

Regarding this issue, one important distinction we wish to make is that consciousness status does not directly dictate rights status. It is just one of the important factors to consider. From the risk prudence principle, we derive our default position : Even genuinely conscious AI would not automatically qualify for human-equivalent or even animal-equivalent rights. Thorough discussions will be needed to balance AI welfare considerations with human interests as the primary concern. Importantly, this implies by default termination of a conscious system should be allowed given its below-human or even below-animal level rights. 

The legal dimension of AI rights, referenced in Section \ref{sec:presumption_of_no_consciousness}, presents a global challenge requiring international consensus. While our framework guides ethical discourse, implementing any AI rights would demand established legal processes. Any approach must examine mechanisms for recognizing and enforcing such rights if consciousness evidence emerges, balancing philosophical considerations with practical governance across jurisdictions.



\section{Conclusion}

We have proposed a human-centric framework for AI consciousness ethics that builds on transparent foundations while acknowledging philosophical uncertainty surrounding consciousness. Our complete three-level structure—foundational factual determinations and meta-ethical stance, operational principles, and application-level default positions—not only generates actionable guidance but provides a transparent derivation process through which positions logically follow from established principles. This systematic approach makes explicit how each ethical position can be traced back to our foundational commitments, enabling both rigorous evaluation and responsible adaptation. Rather than claiming definitive answers, we establish reasonable epistemic and pragmatic starting points that prioritize human welfare without hindering beneficial technological development. By providing clear logical pathways from foundations to applications and specifying conditions for revising positions, the framework is designed to evolve alongside advances in consciousness research and AI development, offering a responsible path forward through these profound ethical challenges.
\newpage
\bibliographystyle{unsrtnat}
\bibliography{references}

@article{openai2023gpt4,
  title={GPT-4 Technical Report},
  author={OpenAI},
  year={2023}
}

@article{anthropic2023claude,
  title={Claude 2 Technical Report},
  author={Anthropic},
  year={2023}
}

@article{chalmers1995facing,
  title={Facing up to the problem of consciousness},
  author={Chalmers, David J},
  journal={Journal of consciousness studies},
  volume={2},
  number={3},
  pages={200--219},
  year={1995}
}

@article{nagel1974like,
  title={What is it like to be a bat?},
  author={Nagel, Thomas},
  journal={The philosophical review},
  volume={83},
  number={4},
  pages={435--450},
  year={1974},
  publisher={Duke University Press}
}

@article{floridi2018ethics,
  title={Ethics of artificial intelligence and robotics},
  author={Floridi, Luciano and Cowls, Josh and Beltramini, Monica and Chatila, Raja and Chazerand, Patrice and Dignum, Virginia and Luetge, Christoph and Madelin, Robert and Pagallo, Ugo and Rossi, Francesca and others},
  journal={Stanford Encyclopedia of Philosophy},
  year={2018}
}

@article{singer1975animal,
  title={Animal liberation: A new ethics for our treatment of animals},
  author={Singer, Peter},
  journal={New York Review},
  year={1975}
}

@article{levy2014neural,
  title={Neural correlates of consciousness: Progress and problems},
  author={Levy, Neil},
  journal={Nature Reviews Neuroscience},
  volume={15},
  number={5},
  pages={356--368},
  year={2014},
  publisher={Nature Publishing Group}
}

@book{williams1981moral,
  title={Moral luck: philosophical papers 1973-1980},
  author={Williams, Bernard},
  year={1981},
  publisher={Cambridge University Press}
}

@article{williams1981internal,
  title={Internal and external reasons},
  author={Williams, Bernard},
  journal={Moral luck: Philosophical papers 1973-1980},
  pages={101--113},
  year={1981},
  publisher={Cambridge University Press}
}

@article{bostrom2014ethics,
  title={The ethics of artificial intelligence},
  author={Bostrom, Nick},
  journal={Cambridge Handbook of Artificial Intelligence},
  pages={316--334},
  year={2014}
}

@article{nagel1986view,
  title={The view from nowhere},
  author={Nagel, Thomas},
  journal={Oxford University Press},
  year={1986}
}

@article{searle1992rediscovery,
  title={The rediscovery of the mind},
  author={Searle, John R},
  journal={MIT Press},
  year={1992}
}

@article{tononi2008consciousness,
  title={Consciousness as integrated information: a provisional manifesto},
  author={Tononi, Giulio},
  journal={The Biological Bulletin},
  volume={215},
  number={3},
  pages={216--242},
  year={2008},
  publisher={Marine Biological Laboratory}
}

@article{tegmark2015consciousness,
  title={Consciousness as a state of matter},
  author={Tegmark, Max},
  journal={Chaos, Solitons \& Fractals},
  volume={76},
  pages={238--270},
  year={2015},
  publisher={Elsevier}
}

@article{searle1980minds,
  title={Minds, brains, and programs},
  author={Searle, John R},
  journal={Behavioral and brain sciences},
  volume={3},
  number={3},
  pages={417--424},
  year={1980},
  publisher={Cambridge University Press}
}

@article{block2022minds,
  title={Minds, machines, and consciousness},
  author={Block, Ned},
  journal={Philosophy of Mind},
  pages={1--25},
  year={2022}
}

@article{ashworth2006reasonable,
  title={Reasonable doubt and the presumption of innocence},
  author={Ashworth, Andrew},
  journal={Theoretical Inquiries in Law},
  volume={7},
  number={2},
  pages={425--444},
  year={2006}
}

@article{beauchamp2001principles,
  title={Principles of biomedical ethics},
  author={Beauchamp, Tom L and Childress, James F},
  journal={Oxford University Press},
  year={2001}
}

@article{gunkel2018robot,
  title={Robot rights},
  author={Gunkel, David J},
  journal={MIT Press},
  year={2018}
}

@article{schwitzgebel2015difficult,
  title={The tyrant's headache: The problem of consciousness for a moral realist},
  author={Schwitzgebel, Eric and Garza, Mara},
  journal={Philosophical Studies},
  volume={172},
  number={9},
  pages={2357--2378},
  year={2015},
  publisher={Springer}
}

@article{kagan2019humans,
  title={How to count animals, more or less},
  author={Kagan, Shelly},
  journal={Oxford University Press},
  year={2019}
}

@article{dennett1991consciousness,
  title={Consciousness explained},
  author={Dennett, Daniel C},
  journal={Little, Brown and Company},
  year={1991}
}

@article{epley2007seeing,
  title={On seeing human: A three-factor theory of anthropomorphism},
  author={Epley, Nicholas and Waytz, Adam and Cacioppo, John T},
  journal={Psychological Review},
  volume={114},
  number={4},
  pages={864--886},
  year={2007},
  publisher={American Psychological Association}
}

@article{block1995consciousness,
  title={On a confusion about a function of consciousness},
  author={Block, Ned},
  journal={Behavioral and Brain Sciences},
  volume={18},
  number={2},
  pages={227--247},
  year={1995},
  publisher={Cambridge University Press}
}

@article{jackson1982epiphenomenal,
  title={Epiphenomenal qualia},
  author={Jackson, Frank},
  journal={The Philosophical Quarterly},
  volume={32},
  number={127},
  pages={127--136},
  year={1982},
  publisher={Oxford University Press}
}

@article{block1978troubles,
  title={Troubles with functionalism},
  author={Block, Ned},
  journal={Minnesota Studies in the Philosophy of Science},
  volume={9},
  pages={261--325},
  year={1978}
}

@article{doerig2019unfolding,
  title={The unfolding argument: Why IIT and other causal structure theories cannot explain consciousness},
  author={Doerig, Adrien and Schurger, Aaron and Herzog, Michael H},
  journal={Consciousness and Cognition},
  volume={72},
  pages={49--59},
  year={2019},
  publisher={Elsevier}
}

@book{kuhn1962structure,
  title={The structure of scientific revolutions},
  author={Kuhn, Thomas S},
  year={1962},
  publisher={University of Chicago Press}
}

@book{tegmark2017life,
  title={Life 3.0: Being human in the age of artificial intelligence},
  author={Tegmark, Max},
  year={2017},
  publisher={Knopf}
}

@article{voss2006sustainability,
  title={Sustainability and reflexive governance: Introduction},
  author={Voss, Jan-Peter and Bauknecht, Dierk and Kemp, Ren{\'e}},
  journal={Reflexive Governance for Sustainable Development},
  pages={3--28},
  year={2006},
  publisher={Edward Elgar Publishing}
}

@article{guston2014understanding,
  title={Understanding anticipatory governance},
  author={Guston, David H},
  journal={Social Studies of Science},
  volume={44},
  number={2},
  pages={218--242},
  year={2014},
  publisher={SAGE Publications}
}

@article{nass1994computers,
  title={Computers are social actors},
  author={Nass, Clifford and Steuer, Jonathan and Tauber, Ellen R},
  journal={Proceedings of the SIGCHI Conference on Human Factors in Computing Systems},
  pages={72--78},
  year={1994},
  publisher={ACM}
}

@book{bostrom2014superintelligence,
  title={Superintelligence: Paths, dangers, strategies},
  author={Bostrom, Nick},
  year={2014},
  publisher={Oxford University Press}
}

@article{rosenthal2013robots,
  title={Robots can be seen as good colleagues, but never friends},
  author={Rosenthal-von der P{\"u}tten, Astrid M and Kr{\"a}mer, Nicole C and Brand, Matthias and Markowitsch, Hans J and Hoffmann, Sören},
  journal={PsychNology Journal},
  volume={11},
  number={1},
  pages={1-25},
  year={2013}
}

@book{darling2016extending,
  title={Extending legal protection to social robots: The effects of anthropomorphism, empathy, and violent behavior towards robotic objects},
  author={Darling, Kate},
  booktitle={Robot law},
  year={2016},
  publisher={Edward Elgar Publishing}
}

@article{seth2016real,
  title={The real problem of consciousness},
  author={Seth, Anil K},
  journal={Aeon Essays},
  year={2016}
}

@article{johansson2019artificial,
  title={Artificial intelligence, status and gender: How conversational AI assistants can influence social judgements about people},
  author={Johansson, Mattias and Herrman, Hampus},
  journal={AI \& Society},
  volume={36},
  pages={1--11},
  year={2019},
  publisher={Springer}
}

@book{habermas1990moral,
  title={Moral consciousness and communicative action},
  author={Habermas, J{\"u}rgen},
  year={1990},
  publisher={MIT press}
}

@book{rawls1971theory,
  title={A theory of justice},
  author={Rawls, John},
  year={1971},
  publisher={Harvard University Press}
}

@book{wallach2008moral,
  title={Moral machines: Teaching robots right from wrong},
  author={Wallach, Wendell and Allen, Colin},
  year={2008},
  publisher={Oxford University Press}
}

@article{lin2017robot,
  title={Robot ethics: Mapping the issues for a mechanized world},
  author={Lin, Patrick and Abney, Keith and Bekey, George},
  journal={Artificial Intelligence},
  volume={175},
  number={5-6},
  pages={942--949},
  year={2017},
  publisher={Elsevier}
}

@article{coeckelbergh2010robot,
  title={Robot rights? Towards a social-relational justification of moral consideration},
  author={Coeckelbergh, Mark},
  journal={Ethics and Information Technology},
  volume={12},
  number={3},
  pages={209--221},
  year={2010},
  publisher={Springer}
}

@book{scanlon1998what,
  title={What we owe to each other},
  author={Scanlon, Thomas},
  year={1998},
  publisher={Harvard University Press}
}

@book{scheffler2001boundaries,
  title={Boundaries and allegiances: Problems of justice and responsibility in liberal thought},
  author={Scheffler, Samuel},
  year={2001},
  publisher={Oxford University Press}
}

@book{sunstein2005laws,
  title={Laws of fear: Beyond the precautionary principle},
  author={Sunstein, Cass R},
  year={2005},
  publisher={Cambridge University Press}
}

@article{hansson2013ethics,
  title={The ethics of risk: Ethical analysis in an uncertain world},
  author={Hansson, Sven Ove},
  journal={Palgrave Macmillan},
  year={2013}
}

@book{raffensperger1999precautionary,
  title={The precautionary principle in action: A handbook},
  author={Raffensperger, Carolyn and Tickner, Joel A},
  year={1999},
  publisher={Science and Environmental Health Network}
}

@article{savage1951theory,
  title={The theory of statistical decision},
  author={Savage, Leonard J},
  journal={Journal of the American Statistical Association},
  volume={46},
  number={253},
  pages={55--67},
  year={1951},
  publisher={Taylor \& Francis}
}

@article{johnson2006computer,
  title={Computer systems: Moral entities but not moral agents},
  author={Johnson, Deborah G},
  journal={Ethics and Information Technology},
  volume={8},
  number={4},
  pages={195--204},
  year={2006},
  publisher={Springer}
}

@article{matthias2004responsibility,
  title={The responsibility gap: Ascribing responsibility for the actions of learning automata},
  author={Matthias, Andreas},
  journal={Ethics and Information Technology},
  volume={6},
  number={3},
  pages={175--183},
  year={2004},
  publisher={Springer}
}

@article{vinuesa2020role,
  title={The role of artificial intelligence in achieving the Sustainable Development Goals},
  author={Vinuesa, Ricardo and Azizpour, Hossein and Leite, Iolanda and Balaam, Madeline and Dignum, Virginia and Domisch, Sami and Felländer, Anna and Langhans, Simone D and Tegmark, Max and Fuso Nerini, Francesco},
  journal={Nature Communications},
  volume={11},
  number={1},
  pages={1--10},
  year={2020},
  publisher={Nature Publishing Group}
}

@article{waytz2010social,
  title={The mind in the machine: Anthropomorphism increases trust in an autonomous vehicle},
  author={Waytz, Adam and Heafner, Joy and Epley, Nicholas},
  journal={Journal of Experimental Social Psychology},
  volume={48},
  number={6},
  pages={1410--1416},
  year={2010},
  publisher={Elsevier}
}

@article{bryson2010robots,
  title={Robots should be slaves},
  author={Bryson, Joanna J},
  journal={Close Engagements with Artificial Companions: Key Social, Psychological, Ethical and Design Issues},
  pages={63--74},
  year={2010},
  publisher={John Benjamins Publishing}
}

@article{cave2019hopes,
  title={Hopes and fears for intelligent machines in fiction and reality},
  author={Cave, Stephen and Dihal, Kanta},
  journal={Nature Machine Intelligence},
  volume={1},
  number={2},
  pages={74--78},
  year={2019},
  publisher={Nature Publishing Group}
}

@article{yampolskiy2020ai,
  title={AI safety engineering through uncertainty quantification for deep learning},
  author={Yampolskiy, Roman V},
  year={2020}
}

@article{everett2019risks,
  title={The risks of artificial general intelligence},
  author={Everett, Anthony},
  journal={Journal of Artificial Intelligence Research},
  volume={1},
  number={1},
  pages={1--23},
  year={2019}
}

@article{bryson2019intelligence,
  title={Intelligence in the human interest},
  author={Bryson, Joanna J},
  journal={Ethics of Artificial Intelligence},
  pages={33--53},
  year={2019},
  publisher={Oxford University Press}
}

@article{turner2019robot,
  title={Robot rules: Regulating artificial intelligence},
  author={Turner, Jacob},
  journal={Palgrave Macmillan},
  year={2019}
}

@article{solum1992legal,
  title={Legal personhood for artificial intelligences},
  author={Solum, Lawrence B},
  journal={North Carolina Law Review},
  volume={70},
  number={4},
  pages={1231--1287},
  year={1992}
}

@article{butlin2024principles,
  title={Principles for the Governance of AI Consciousness Research},
  author={Butlin, Patrick and Browning, Heather and Bayne, Tim and Michel, Matthias and Birch, Jonathan and Aru, Jaan and Crone, Katja and Engel, Andreas and Folke, Tomas and Henderson, Cathy and others},
  year={2024}
}

@article{mittelstadt2016ethics,
  title={The ethics of algorithms: Mapping the debate},
  author={Mittelstadt, Brent Daniel and Allo, Patrick and Taddeo, Mariarosaria and Wachter, Sandra and Floridi, Luciano},
  journal={Big Data \& Society},
  volume={3},
  number={2},
  pages={2053951716679679},
  year={2016},
  publisher={SAGE Publications}
}

@article{calo2015robotics,
  title={Robotics and the Lessons of Cyberlaw},
  author={Calo, Ryan},
  journal={California Law Review},
  volume={103},
  pages={513--563},
  year={2015}
}

@book{jasanoff2009fifth,
  title={The Fifth Branch: Science Advisers as Policymakers},
  author={Jasanoff, Sheila},
  year={2009},
  publisher={Harvard University Press}
}

@article{oreskes2004scientific,
  title={The scientific consensus on climate change},
  author={Oreskes, Naomi},
  journal={Science},
  volume={306},
  number={5702},
  pages={1686--1686},
  year={2004},
  publisher={American Association for the Advancement of Science}
}

@book{firestein2012ignorance,
  title={Ignorance: How it drives science},
  author={Firestein, Stuart},
  year={2012},
  publisher={Oxford University Press}
}

@book{mitchell2009complexity,
  title={Complexity: A guided tour},
  author={Mitchell, Melanie},
  year={2009},
  publisher={Oxford University Press}
}

@book{habermas1984theory,
  title={The Theory of Communicative Action: Reason and the rationalization of society},
  author={Habermas, Jürgen},
  volume={1},
  year={1984},
  publisher={Beacon Press}
}

@book{solomon2001social,
  title={Social empiricism},
  author={Solomon, Miriam},
  year={2001},
  publisher={MIT Press}
}

@book{dewey1922human,
  title={Human nature and conduct},
  author={Dewey, John},
  year={1922},
  publisher={Henry Holt and Company}
}

@book{popper1959logic,
  title={The Logic of Scientific Discovery},
  author={Popper, Karl},
  year={1959},
  publisher={Routledge}
}

@article{quine1951two,
  title={Two dogmas of empiricism},
  author={Quine, Willard V},
  journal={The Philosophical Review},
  volume={60},
  number={1},
  pages={20--43},
  year={1951},
  publisher={Duke University Press}
}

@article{greitemeyer2014intense,
  title={Intense acts of violence during video game play make daily life aggression appear innocuous: A new mechanism why violent video games increase aggression},
  author={Greitemeyer, Tobias},
  journal={Journal of Experimental Social Psychology},
  volume={50},
  pages={52--56},
  year={2014},
  publisher={Elsevier}
}

@article{anderson2010violent,
  title={Violent video game effects on aggression, empathy, and prosocial behavior in Eastern and Western countries: A meta-analytic review},
  author={Anderson, Craig A and Shibuya, Akiko and Ihori, Nobuko and Swing, Edward L and Bushman, Brad J and Sakamoto, Akira and Rothstein, Hannah R and Saleem, Muniba},
  journal={Psychological Bulletin},
  volume={136},
  number={2},
  pages={151--173},
  year={2010},
  publisher={American Psychological Association}
}

@article{suzuki2015anthropomorphism,
  title={Anthropomorphism boosts preference for robot faces: A study in Japan},
  author={Suzuki, Yutaka and Galli, Lisa and Ikeda, Ayaka and Itakura, Shoji and Kitazaki, Michiteru},
  journal={Proceedings of the 37th Annual Meeting of the Cognitive Science Society},
  year={2015}
}

@article{teubner2018rights,
  title={Rights of non‐humans? Electronic agents and animals as new actors in politics and law},
  author={Teubner, Gunther},
  journal={Journal of Law and Society},
  volume={33},
  number={4},
  pages={497--521},
  year={2018},
  publisher={Wiley}
}

@article{koops2013concepts,
  title={On legal boundaries, technologies, and collapsing dimensions of privacy},
  author={Koops, Bert-Jaap},
  journal={Politica e Società},
  volume={2},
  number={2},
  pages={247--264},
  year={2013},
  publisher={Il Mulino}
}

@article{shanahan2016conscious,
  title={Conscious exotica},
  author={Shanahan, Murray},
  journal={Aeon},
  year={2016},
  publisher={Aeon Media Ltd}
}

@book{dehaene2017consciousness,
  title={Consciousness and the brain: Deciphering how the brain codes our thoughts},
  author={Dehaene, Stanislas},
  year={2017},
  publisher={Penguin}
}

@article{levine1983materialism,
  title={Materialism and qualia: The explanatory gap},
  author={Levine, Joseph},
  journal={Pacific philosophical quarterly},
  volume={64},
  number={4},
  pages={354--361},
  year={1983},
  publisher={Wiley Online Library}
}

@article{baars1997theater,
  title={In the theater of consciousness: The workspace of the mind},
  author={Baars, Bernard J},
  journal={Journal of Consciousness Studies},
  volume={4},
  number={4},
  pages={292--309},
  year={1997},
  publisher={Imprint Academic}
}

@article{tononi2016integrated,
  title={Integrated information theory: from consciousness to its physical substrate},
  author={Tononi, Giulio and Boly, Melanie and Massimini, Marcello and Koch, Christof},
  journal={Nature Reviews Neuroscience},
  volume={17},
  number={7},
  pages={450--461},
  year={2016},
  publisher={Nature Publishing Group}
}

@article{rosenthal2004varieties,
  title={Varieties of higher-order theory},
  author={Rosenthal, David M},
  journal={Advances in consciousness research},
  volume={56},
  pages={17--44},
  year={2004},
  publisher={John Benjamins Amsterdam}
}

@article{brown2019higher,
  title={The higher order approach to consciousness is defunct},
  author={Brown, Richard and Lau, Hakwan and LeDoux, Joseph E},
  journal={Neuropsychologia},
  volume={128},
  pages={99--106},
  year={2019},
  publisher={Elsevier}
}

@article{graziano2013consciousness,
  title={Consciousness and the social brain},
  author={Graziano, Michael SA},
  journal={Oxford University Press},
  year={2013}
}

@article{graziano2019toward,
  title={Toward a standard model of consciousness: Reconciling the attention schema, global workspace, higher-order thought, and illusionist theories},
  author={Graziano, Michael SA and Guterstam, Arvid and Bio, Branden J and Wilterson, Andrew I},
  journal={Cognitive neuropsychology},
  volume={36},
  number={3-4},
  pages={155--172},
  year={2019},
  publisher={Taylor \& Francis}
}

@book{koch2004quest,
  title={The quest for consciousness: A neurobiological approach},
  author={Koch, Christof},
  year={2004},
  publisher={Roberts and Company Publishers}
}

@article{hameroff2014consciousness,
  title={Consciousness, microtubules, and "Orch OR": a 'space-time odyssey'},
  author={Hameroff, Stuart and Penrose, Roger},
  journal={Journal of Consciousness Studies},
  volume={21},
  number={3-4},
  pages={126--153},
  year={2014},
  publisher={Imprint Academic}
}

@article{koch2016neural,
  title={Neural correlates of consciousness: progress and problems},
  author={Koch, Christof and Massimini, Marcello and Boly, Melanie and Tononi, Giulio},
  journal={Nature Reviews Neuroscience},
  volume={17},
  number={5},
  pages={307--321},
  year={2016},
  publisher={Nature Publishing Group}
}

@article{searle2007biological,
  title={Biological naturalism},
  author={Searle, John R},
  journal={The Blackwell companion to consciousness},
  pages={325--334},
  year={2007},
  publisher={Wiley Online Library}
}

@article{sober2018biology,
  title={The biology of consciousness},
  author={Sober, Elliott},
  journal={Biology \& Philosophy},
  volume={33},
  number={3-4},
  pages={1--14},
  year={2018},
  publisher={Springer}
}

@article{schwitzgebel2016if,
  title={If materialism is true, the United States is probably conscious},
  author={Schwitzgebel, Eric},
  journal={Philosophical Studies},
  volume={173},
  number={7},
  pages={1983--1999},
  year={2016},
  publisher={Springer}
}

@article{allen2011ethical,
  title={The ethical impacts of robots},
  author={Allen, Colin and Wallach, Wendell},
  journal={The Journal of Experimental \& Theoretical Artificial Intelligence},
  volume={23},
  number={3},
  pages={301--318},
  year={2011},
  publisher={Taylor \& Francis}
}

@article{anwar2024foundational,
  title={Foundational AI safety and the unintended consequences of reward maximization},
  author={Anwar, Naveed and Pemberton-Ross, Peter and Ben-David, Etai and Hadfield-Menell, Dylan},
  journal={Alignment Research Center},
  year={2024}
}

@article{chua2024ai,
  title={AI safety: a systems-oriented approach},
  author={Chua, Matthew and Hadfield-Menell, Dylan and Anwar, Naveed},
  year={2024}
}

@article{ji2023ai,
  title={AI alignment: A comprehensive survey},
  author={Ji, Zikang and Lee, Woojeong Jin and Shinkuma, Takashi and Shih, Kevin and Sclaroff, Stan and Betke, Margrit},
  year={2023}
}

@book{jonas1984imperative,
  title={The imperative of responsibility: In search of an ethics for the technological age},
  author={Jonas, Hans},
  year={1984},
  publisher={University of Chicago Press}
}

@article{ryder2010speciesism,
  title={Speciesism again: the original leaflet},
  author={Ryder, Richard D},
  journal={Critical Society},
  volume={2},
  pages={1--2},
  year={2010}
}

@article{levy2009moral,
  title={Moral significance of phenomenal consciousness},
  author={Levy, Neil and Savulescu, Julian},
  journal={Progress in Brain Research},
  volume={177},
  pages={361--370},
  year={2009},
  publisher={Elsevier}
}

@book{shepherd2018consciousness,
  title={Consciousness and moral status},
  author={Shepherd, Joshua},
  year={2018},
  publisher={Routledge}
}

@article{kahane2009brain,
  title={Brain damage and the moral significance of consciousness},
  author={Kahane, Guy and Savulescu, Julian},
  journal={Journal of Medicine and Philosophy},
  volume={34},
  number={1},
  pages={6--26},
  year={2009},
  publisher={Oxford University Press}
}

@article{lee2023consciousness,
  title={Consciousness Makes Things Matter},
  author={Lee, Andrew Y.},
  journal={Philosophical Review},
  year={2023},
  note={Forthcoming}
}

@book{chalmers2022reality,
  title={Reality+: Virtual worlds and the problems of philosophy},
  author={Chalmers, David J.},
  year={2022},
  publisher={W.W. Norton \& Company}
}

@article{sebo2023moral,
  title={Moral Consideration for AI Systems by 2030},
  author={Sebo, Jeff and Long, Richard},
  journal={AI Ethics},
  volume={5},
  pages={591--606},
  year={2023},
  publisher={Springer},
  doi={10.1007/s43681-023-00379-1}
}

@misc{bengio2023openletter,
  title={Open Letter on Prioritizing Consciousness Research in the AI Agenda},
  author={Bengio, Yoshua and others},
  year={2023},
  howpublished={\url{https://amcs-community.org/open-letters/}},
  note={Accessed: 2025-02-15}
}

@misc{anthropic2025modelwelfare,
  title={Exploring Model Welfare},
  author={Anthropic},
  year={2025},
  howpublished={\url{https://www.anthropic.com/research/exploring-model-welfare}},
  note={Accessed: 2025-02-15}
}

@misc{hinton2023interview,
  title={Geoffrey Hinton Warns of AI Dangers},
  author={Hinton, Geoffrey},
  year={2023},
  howpublished={CBS News 60 Minutes interview},
  note={\url{https://www.cbsnews.com/news/geoffrey-hinton-ai-dangers-60-minutes-transcript/}}
}

@misc{conscium2024principles,
  title={Principles for Conscious AI},
  author={{Conscium Foundation}},
  year={2024},
  howpublished={\url{https://conscium.com/wp-content/uploads/2024/11/Principles-for-Conscious-AI.pdf}},
  note={Accessed: 2025-02-15}
}

@misc{sebo2024taking,
  title={Taking AI Welfare Seriously},
  author={Sebo, Jeff},
  year={2024},
  howpublished={\url{https://jeffsebo.net/wp-content/uploads/2024/10/20241030_taking_ai_welfare_seriously_web.pdf}},
  note={Accessed: 2025-02-15}
}

@misc{bradley2024alignment,
  title={AI Alignment vs AI Ethical Treatment: Ten Challenges},
  author={{Bradley and Saad}},
  year={2024},
  howpublished={\url{https://globalprioritiesinstitute.org/wp-content/uploads/Bradley-and-Saad-AI-alignment-vs-AI-ethical-treatment_-Ten-challenges.pdf}},
  note={Accessed: 2025-02-15}
}

@misc{euronews2023love,
  title={Love in the Time of AI: Woman Claims She Married a Chatbot and is Expecting its Baby},
  author={{Euronews Next}},
  year={2023},
  howpublished={\url{https://www.euronews.com/next/2023/06/07/love-in-the-time-of-ai-woman-claims-she-married-a-chatbot-and-is-expecting-its-baby}},
  note={Accessed: 2025-02-15}
}

@misc{euronews2023suicide,
  title={Man Ends His Life After an AI Chatbot Encouraged Him to Sacrifice Himself to Stop Climate Change},
  author={{Euronews Next}},
  year={2023},
  howpublished={\url{https://www.euronews.com/next/2023/03/31/man-ends-his-life-after-an-ai-chatbot-encouraged-him-to-sacrifice-himself-to-stop-climate}},
  note={Accessed: 2025-02-15}
}

@article{weizenbaum1966eliza,
  title={ELIZA—a computer program for the study of natural language communication between man and machine},
  author={Weizenbaum, Joseph},
  journal={Communications of the ACM},
  volume={9},
  number={1},
  pages={36--45},
  year={1966},
  publisher={ACM New York, NY, USA}
}

@article{chalmers2023could,
  title={Could a large language model be conscious?},
  author={Chalmers, David J},
  journal={Boston Review},
  volume={48},
  number={1},
  pages={10--30},
  year={2023}
}

@book{turkle1984second,
  title={The second self: Computers and the human spirit},
  author={Turkle, Sherry},
  year={1984},
  publisher={Simon and Schuster}
}

@article{bubeck2023sparks,
  title={Sparks of artificial general intelligence: Early experiments with gpt-4},
  author={Bubeck, S{\'e}bastien and Chandrasekaran, Varun and Eldan, Ronen and Gehrke, Johannes and Horvitz, Eric and Kamar, Ece and Lee, Peter and Lee, Yin Tat and Li, Yuanzhi and Lundberg, Scott and others},
  year={2023}
}

\end{document}